\documentclass[english,aps,pra,showpacs,twocolumn,nofootinbib]{revtex4}
\usepackage[T1]{fontenc}
\usepackage[latin9]{inputenc}
\usepackage{color}
\usepackage{amsmath}
\usepackage{graphicx}
\usepackage{amssymb}

\newcommand{\lyxdot}{.}

\begin{document}

\title{Dynamics of Morphology-Dependent Resonances by Openness in Dielectric
Disk for TE polarization}

\author{Jinhang Cho}

\affiliation{Acceleration Research Center for Quantum Chaos Applications, Sogang
University, Seoul 121-742, Korea}

\author{Sunghwan Rim}

\affiliation{Acceleration Research Center for Quantum Chaos Applications, Sogang
University, Seoul 121-742, Korea}

\author{Chil-Min Kim}

\affiliation{Acceleration Research Center for Quantum Chaos Applications, Sogang
University, Seoul 121-742, Korea}

\author{}
\begin{abstract}
 We have studied the parametric-evolution of morphology-dependent
resonances according to the change of openness in a two-dimensional
dielectric microdisk for TE polarization. For the first time, we report
that the dynamics exhibits avoided resonance crossings between inner
and outer resonances  even though the corresponding billiard is integrable.
 Due to these recondite avoidances, inner and outer resonances can
be exchanged and $Q$-factor of inner resonances is strongly affected.
We analyze the diverse phenomena arisen from the dynamics including
the avoided crossings.
\end{abstract}



\maketitle

\section{Introduction}

Morphology-dependent resonances (MDRs), which is defined as the resonances
found in cylindrical, spherical and ellipsoidal shaped optical cavities,
are a much studied topic in the field of optics due to their practical
importance in many applications \cite{Chang} as well as their heuristic
value. More than a decade ago, Johnson is the first to approach MDR
in a spherical cavity \cite{Johnson} as an analogy of quantum mechanical
resonances, which is familiar in nuclear, atomic, and molecular scattering
theory. By the same approach, N{\"o}ckel have studied MDR analytically
in a cylindrical cavity \cite{Noeckel} as an unperturbed system in
order to analyze a slightly deformed one.

Recently, study of the MDR is resurfaced as a hot issue in connection
with the application of deformed microcavity lasers because of its
high potentiality of application for photonics and optoelectronics.
Bogomolny \textit{et al.} have reported the existence of a different
set of resonances, which can be classified as shape (\textit{outer})
resonances (\textit{ORs}) with high leakage along with Feshbach (\textit{inner})
resonances (\textit{IRs}) in a two-dimensional dielectric circular
microdisk for transverse magnetic (TM) and transverse electric (TE)
polarizations \cite{Dubertrand,Bogomolny}. The similar topic was
treated in a numerical method for quantum billiard \cite{Harayama}.
The evolutions of these two classes of resonances to the so-called
\textit{small opening limit} where the refractive index $n$ of the
disk goes to infinity have been explicitly studied \cite{Dettmann}
and the applicability of effective potential analogy for the ORs has
been checked \cite{Cho}. One intriguing fact of the MDR dynamics
from last two studies is that, for TE polarization with magnetic
field $\vec{H}$ perpendicular to the cavity plane, both IRs and ORs
are found in the same region of complex wavenumber plane  where
the regime of $n$ is practically relevant. 

In two-dimensional chaotic quantum billiards, it is well known that
the level dynamics resembles dynamics of one-dimensional particles
with repulsive interaction as a system parameter varies \cite{Stoeckmann}.
It is called avoided level crossing. Similarly, avoided resonance
crossing (ARC) phenomenon in open Hamiltonian systems is generally
categorized into two classes \cite{ARC_open_Hamiltonian}. In the
case of a strong interaction, shortly \textit{strong ARC}, the imaginary
parts of complex-valued energies are crossing and the real parts
are avoided. The identities of the two modes are exchanged via the
ARC and their wave patterns are strongly superposed during the ARC.
For a weak interaction, shortly \textit{weak ARC}, the real parts
are crossing and the imaginary parts are avoided while their identities
are retained.

In two-dimensional chaotic dielectric cavities, there have been undertaken
many theoretical \cite{ARC_chaotic_theory_Wiersig_Hentschel,ARC_chaotic_theory_SYLee,ARC_chaotic_theory_Wiersig_SWKim_Hentschel,ARC_chaotic_theory_Ryu,ARC_chaotic_theory_Lizuain,ARC_chaotic_theory_Poli}
and experimental \cite{ARC_chaotic_exp_Dietz,ARC_chaotic_exp_SBLee,ARC_chaotic_exp_SBLee2}
researches related with the ARC.  Due to the open property, the resonance
dynamics in a dielectric cavity is much complicated and shows diverse
aspects compared to that of the closed billiard. Recently, it has
been represented that, when a deformation parameter is varied in the
rectangular or elliptic cavity, of which the corresponding billiard
is integrable, ARC can occur by the effect of openness \cite{ARC_integrable_Wiersig,ARC_integrable_Unterhinninghofen}.%
\footnote{In the pure one-dimensional system like a double-well potential, integrability
does not forbid avoided crossing between two states.%
}   Another recent report have claimed  that, even on a slight deformation
from the disk shape, a symmetry breaking of escaping rays occurs,
which gives rise to dramatic changes in emission patterns depending
on the characteristics of deformations \cite{Creagh}, and emission
directions and resonance patterns are deeply influenced by the refractive
index of the cavity, i.e., openness of the system, as well as the
underlying classical manifold structures \cite{Lee}. These imply
that analyzing the emission directions and mode patterns of slightly
deformed cavities requires understanding of the MDR dynamics for openness.

In this paper, we uncover the   mechanism behind the MDR dynamics
in a circular disk for TE polarization as the openness parameter $\Omega=1/n$
is varied and observe the impacts of pure openness on  the dynamics.
The perfect rotational symmetry provides the analysis possible.

\section{Avoided Resonance Crossings in Dielectric Disk for TE polarization}

The generic two-dimensional dielectric cavities have mainly two types
of changeable system parameters, e.g., degree of geometrical deformation
and openness. In such cavities, IRs can generally interact with
each other.  Their resonance dynamics contains a complicated mixture
of the aspect by deformation and the effect of openness. On the other
hand, the circular disk does not have any deformation parameter and
the only relevant parameter is the openness.  Moreover, its rotational
symmetry makes that the angular component of the wavefunction is trivial.
As a result, it renders the system effectively one-dimensional and
then leads more clear and basic study for the pure openness without
the aspect by deformation.

\begin{figure}
\begin{centering}
\includegraphics[width=8.5cm]{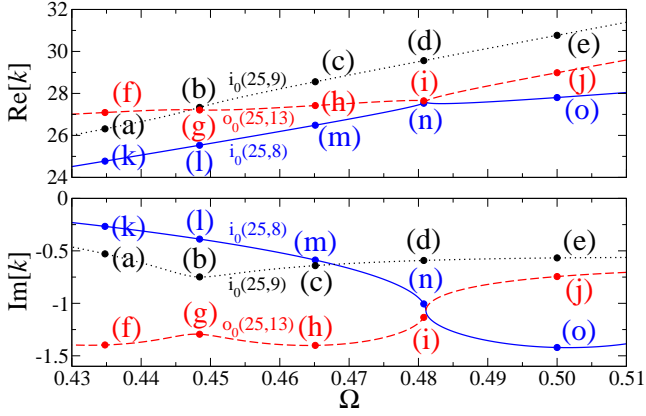}
\par\end{centering}

\caption{(Color online) ARCs between IR and low-leaky OR. Solid, dotted,
and dashed lines are $\text{i}_{0}(25,8)$, $\text{i}_{0}(25,9)$,
and $\text{o}_{0}(25,13)$, respectively. \label{fig:ARCs_disk}}

\end{figure}

The Helmholtz equation of dielectric cavities is given by $\nabla^{2}\psi+\left(k/\Omega\right)^{2}\psi=0$,
where $\psi$ and $k$ are the wavefunction and the wavenumber, respectively.
In the disk with the radius $R=1$, the complex-valued wavenumbers
$k_{r}$ for resonances are obtained from the boundary matching conditions
for TM and TE polarizations under the purely outgoing wave condition
\cite{thesis_Noeckel,thesis_Hentschel}.  In the opening regime
$0<\Omega<1$, the IRs, $\text{i}(m,l)$, are  labelled by angular
momentum quantum number $m$ and  modal ordering number $l$ which
corresponds to the number of radial intensity peaks inside the cavity.
Similarly, the ORs can be also represented by $\text{o}(m,l)$, where
$l$ is assigned in accordance with the increasing order of $\text{Re}[k_{r}]$.
For a given $m$, the maximum numbers of ORs in TM and TE polarizations
are $\left[\frac{m}{2}\right]$ and $\left[\frac{m+1}{2}\right]$,
respectively (the notation $\left[\,\right]$ is integer value). 
 Notice that, in TE case, some ORs have relatively low leakage and
we have especially called them \textit{low-leaky outer resonances}
(low-leaky ORs) \cite{Cho}. Since there is only one low-leaky OR
 having the largest $\text{Re}[k_{r}]$ among the ORs for each $m(>0)$,
low-leaky ORs can be denoted by $\text{o}(m,\left[\frac{m+1}{2}\right])$.
  Additionally, the notations $\text{i}_{0}(m,l)$ and $\text{o}_{0}(m,l)$
are used for IR and OR in the small opening limit, respectively, where
they can be unambiguously identified \cite{Cho,Dettmann}. The subscript
zero means $\Omega\to0$.

\begin{figure}
\begin{centering}
\includegraphics[width=8.5cm]{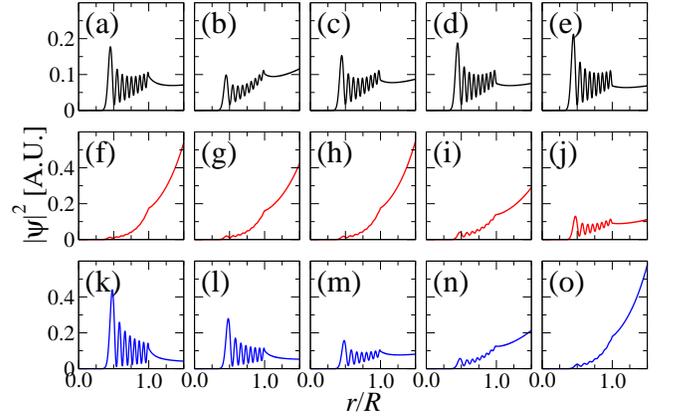}
\par\end{centering}

\caption{(Color online) Metamorphosis of two IRs and a low-leaky OR. Top,
middle, and bottom rows are $\text{i}_{0}(25,9)$, $\text{o}_{0}(25,13)$,
and $\text{i}_{0}(25,8)$, respectively. Five pictures in each row
are at $\Omega=0.435$, $0.448$, $0.465$, $0.481$, and $0.500$.\label{fig:radial_wf}}

\end{figure}

When $\text{o}_{0}(m,\left[\frac{m+1}{2}\right])$ is evolved from
the small opening limit toward the large opening regime,  both
weak ARC and strong ARC are unexpectedly observed in the dielectric
disk. Figure \ref{fig:ARCs_disk} shows the two types of ARCs between
IR and low-leaky OR with $m=25$.  At $\Omega\simeq0.448$, $\text{o}_{0}(25,13)$
interacts with $\text{i}_{0}(25,9)$ via weak ARC. When the openness
increases, $\text{o}_{0}(25,13)$ interacts with $\text{i}_{0}(25,8)$
via strong ARC at $\Omega\simeq0.481$. The radial intensity patterns
at five points on each line in Fig. \ref{fig:ARCs_disk} are shown
in Fig. \ref{fig:radial_wf}. Due to the superposition near the point
of the weak ARC at $\Omega\simeq0.448$, the leakage of $\text{o}(25,13)$
(Fig. \ref{fig:radial_wf}(g)) decreases while that of $\text{i}(25,9)$
(Fig. \ref{fig:radial_wf}(b)) increases. However their identities
are retained after the ARC as shown in Figs. \ref{fig:radial_wf}(c)
and (h). Likewise at the weak ARC, $\text{o}(25,13)$ and $\text{i}(25,8)$
also undergo the change of leakage and their spatial patterns are
superposed near the point of strong ARC as shown in Figs. \ref{fig:radial_wf}(h)
and (m). Especially, just at the point of the strong ARC,  two resonances
are strongly superposed and then their patterns are almost identical
as shown in Figs. \ref{fig:radial_wf}(i) and (n). After the strong
ARC, the identities of $\text{i}(25,8)$ and $\text{o}(25,13)$ are
exchanged as shown in Figs. \ref{fig:radial_wf}(j) and (o). Consequently,
$\text{o}_{0}(25,13)$ becomes $\text{i}(25,8)$ and $\text{i}_{0}(25,8)$
becomes $\text{o}(25,13)$. On a further increase of $\Omega$
over $0.51$, $\text{i}_{0}(25,8)$, which has become $\text{o}(25,13)$,
interacts with $\text{i}_{0}(25,7)$ through another strong ARC. Eventually
$\text{i}_{0}(m,l)$ can experience strong ARCs up to two times 
over the whole regime of $\Omega$. $\text{i}_{0}(m,l)$ changes to
$\text{o}(m,\left[\frac{m+1}{2}\right])$ at the first strong ARC
and  $\text{i}_{0}(m,l)$ to $\text{i}(m,l-1)$ at the following
second strong ARC:

\[
\text{i}_{0}(m,l)\stackrel{_{\text{strong ARC}}}{\longrightarrow}\text{o}(m,\left[\frac{m+1}{2}\right])\stackrel{_{\text{strong ARC}}}{\longrightarrow}\text{i}(m,l-1).\]

On the other hand, $\text{i}_{0}(25,9)$ and $\text{i}_{0}(25,8)$
do not interact each other, even if they have the same symmetry and
their imaginary values of $k_{r}$ cross at $\Omega\simeq0.468$.
It has been known that two-dimensional optical dielectric cavities
are typically non-integrable systems due to their open property and
the openness breaks the integrability of the systems. Consequently,
the avoided crossings between the IRs occur in the optical cavities,
even if their corresponding billiards are integrable. It was explicitly
shown in elliptic and rectangular cavity \cite{ARC_integrable_Wiersig,ARC_integrable_Unterhinninghofen}.
But, unlike the cavities which correspond to integrable billiards
such as rectangular and elliptic cavity, the avoided crossings between
IRs intriguingly do not occur in the case of circular disk cavity
even though the system is open. This fact implicitly indicates that
the IRs remain with the properties of  the normal modes in the corresponding
billiard which is integrable in this case.

\section{Toy Model}

Before we discuss the numerical results, we will consider the analytical
approach through a toy model. The ARCs can be displayed with effective
Hamiltonian composed of a non-Hermitian $2\times2$ matrix and a coupling
term as follows:\begin{equation}
H_{\text{eff}}(\Omega)=\left(\begin{array}{cc}
E_{\text{i}}(\Omega) & 0\\
0 & E_{\text{o}}(\Omega)\end{array}\right)+\left(\begin{array}{cc}
0 & \eta\Omega\\
\xi\Omega & 0\end{array}\right),\label{eq:Heff}\end{equation}
where $E_{\text{i}}(\Omega)$ and $E_{\text{o}}(\Omega)$ are the
complex energy $k_{r}^{2}$ of IR and OR, respectively. The off-diagonal
element in the coupling term, $\eta\in\mathbb{C}$ ($\xi\in\mathbb{C}$),
describes the coupling effect from OR (IR) to IR (OR) and it
is assumed that the coupling effects are proportional to $\Omega$
due to the fact that the coupling must becomes zero in the small opening
limit. Equation \eqref{eq:Heff} leads the eigenvalues \begin{equation}
E_{\pm}(\Omega)=\frac{E_{\text{i}}(\Omega)+E_{\text{o}}(\Omega)}{2}\pm\sqrt{\frac{(E_{\text{i}}(\Omega)-E_{\text{o}}(\Omega))^{2}}{4}+\eta\xi\Omega^{2}}.\label{eq:Heff_eigenvalue}\end{equation}

\begin{figure}
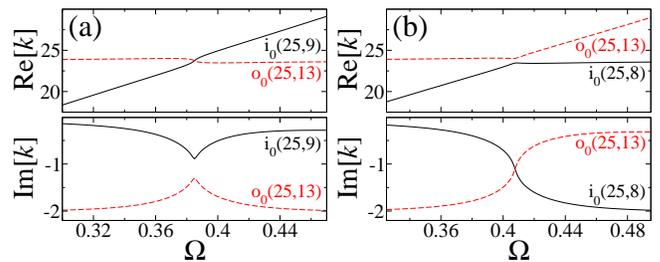

\begin{centering}
\includegraphics[width=4.2cm]{./Fig3a_WARC} \includegraphics[width=4.2cm]{./Fig3b_SARC}
\par\end{centering}

\caption{(Color online) Analytical modeling of (a) weak ARC and (b) strong
ARC. Solid and dashed lines are IR and OR, respectively.\label{fig:ARCs_Heff}}

\end{figure}

To simply illustrate the ARCs, we substitute $E_{\text{i}}(\Omega)$
and $E_{\text{o}}(\Omega)$ with the intuitive relations of the MDR
dynamics. Firstly, to obtain the basic MDR dynamics for TE case except
for the effects of mode interaction and Brewster angle by openness,
we borrow  the  dynamics for TM polarization, in which the interactions
between the two types of resonances are infinitesimal in the experimentally
feasible opening regime due to the absence of low-leaky ORs. From
this, $\text{Re}[k_{r}]$ and $\text{Im}[k_{r}]$ of an IR become
almost linear to $\Omega$ and to $\gamma$, respectively \cite{Noeckel,Dubertrand,Bogomolny}.
Here, $\gamma$ given by \begin{equation}
\gamma=-\frac{\Omega}{2}\ln\frac{1+\Omega}{1-\Omega}\label{eq:Im_bound}\end{equation}
 is the lower bound of  $\text{Im}[k_{r}]$ of IRs in TM case.
 Also an OR in TM case maintains almost an constant complex valued
$k_{r}$ for the change of $\Omega$ \cite{Dubertrand,Bogomolny,Cho}.
Secondly, we consider that the destinations of $nk_{r}$ of IRs and
$k_{r}$ of ORs in the small opening limit for TE case, correspond
to the zeros of Bessel function \cite{Ryu,Dettmann}, $j_{m,l}$,
and the zeros of Hankel function derivatives \cite{Dettmann}, $h_{m,l}^{\prime}$,
respectively. Then, $E_{\text{i}}(\Omega)$ and $E_{\text{o}}(\Omega)$
for TE case can be simply expressed as follows:    \begin{equation}
\begin{array}{c}
E_{\text{i}}(\Omega)\thickapprox\left(j_{m,l}\Omega+i\gamma\right)^{2},\\
E_{\text{o}}(\Omega)\thickapprox\left(h_{m,l}^{\prime}\right)^{2}.\end{array}\label{eq:eigenvalues}\end{equation}
 Even though we borrowed the MDR dynamics for TM case, the dynamics
of $\sqrt{E_{\text{i}}(\Omega)}$ and $\sqrt{E_{\text{o}}(\Omega)}$
was well matched to the numerically obtained resonance tracing behaviors
\cite{Cho} for TE case in the opening regime where the interaction
is very small.  Figures \ref{fig:ARCs_Heff}(a) and \ref{fig:ARCs_Heff}(b)
show weak ARC of $\text{i}_{0}(25,9)$  and strong ARC of $\text{i}_{0}(25,8)$
with $\text{o}_{0}(25,13)$, respectively, when the coupling constant
 $\eta\xi=12737.8$.  The value of coupling constant was properly
chosen as compared with the numerical result Fig. \ref{fig:ARCs_disk}.

 The degree of mode coupling depends on the relative quantity between
$(E_{i}-E_{o})^{2}$ and coupling strength $\eta\xi\Omega^{2}$ located
in the second term of right-hand side of Eq. \eqref{eq:Heff_eigenvalue}.
According to the relative quantity, the crossing point of $\Omega$
are determined and the transition of the type of ARC can be occur.
It is a well-known  phenomenon \cite{ARC_open_Hamiltonian,ARC_chaotic_theory_SYLee,ARC_integrable_Wiersig}.
 In Fig. \ref{fig:ARCs_Heff} obtained from the toy model, (a) is
about $\text{\ensuremath{i_{0}}}(25,9)$ and (b) is about $\text{\ensuremath{i_{0}}}(25,8)$.
Under the our coupling constant matched to the numerical result (Fig.
\ref{fig:ARCs_disk}), the point of transition from weak ARC to strong
ARC for $m=25$ is shown between these two IRs.

\section{Dynamics of Morphology-Dependent Resonances by Openness}

\begin{figure}
\begin{centering}
\includegraphics[width=8.5cm]{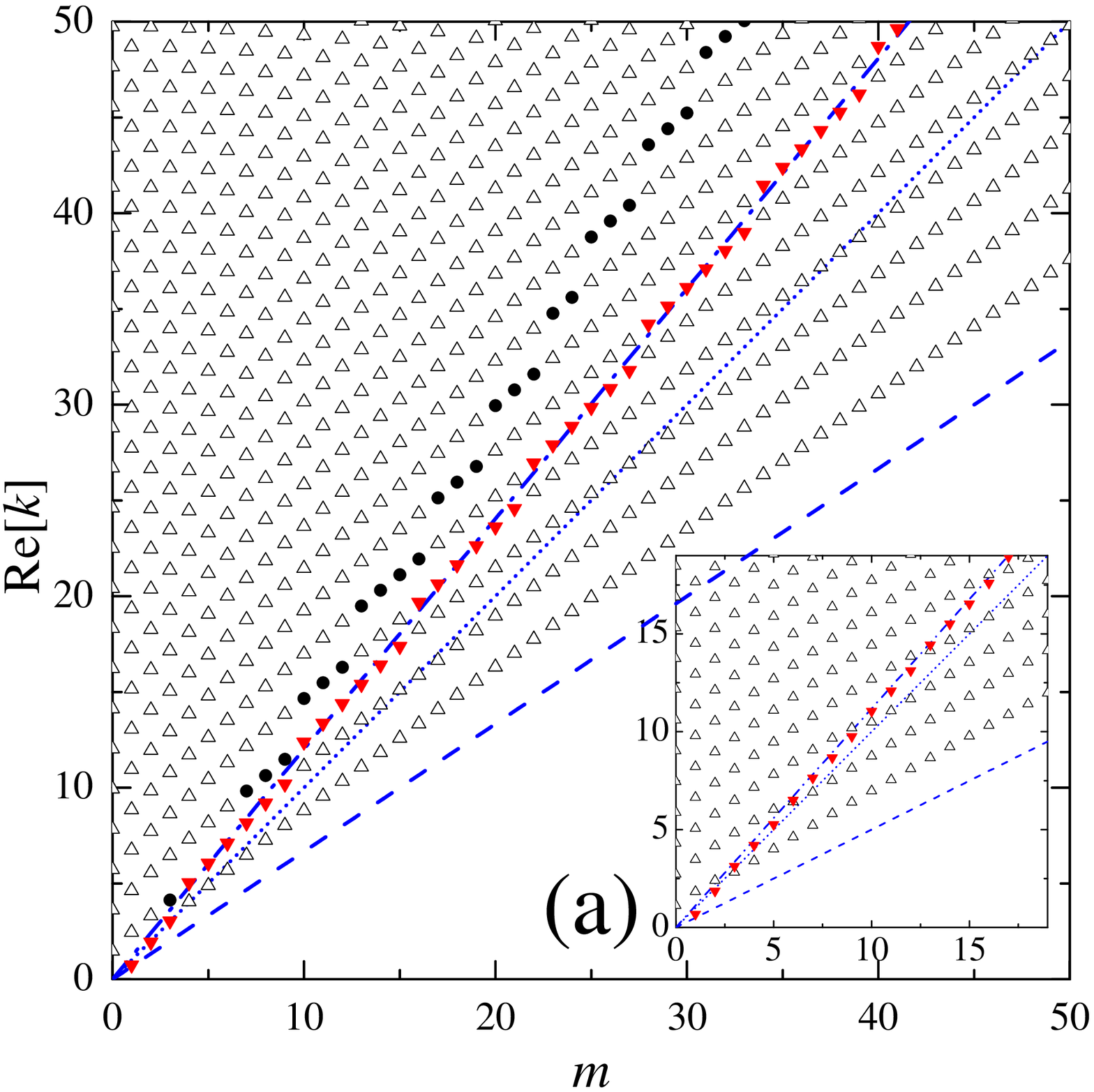}
\par\end{centering}

\begin{centering}
\includegraphics[width=8.5cm]{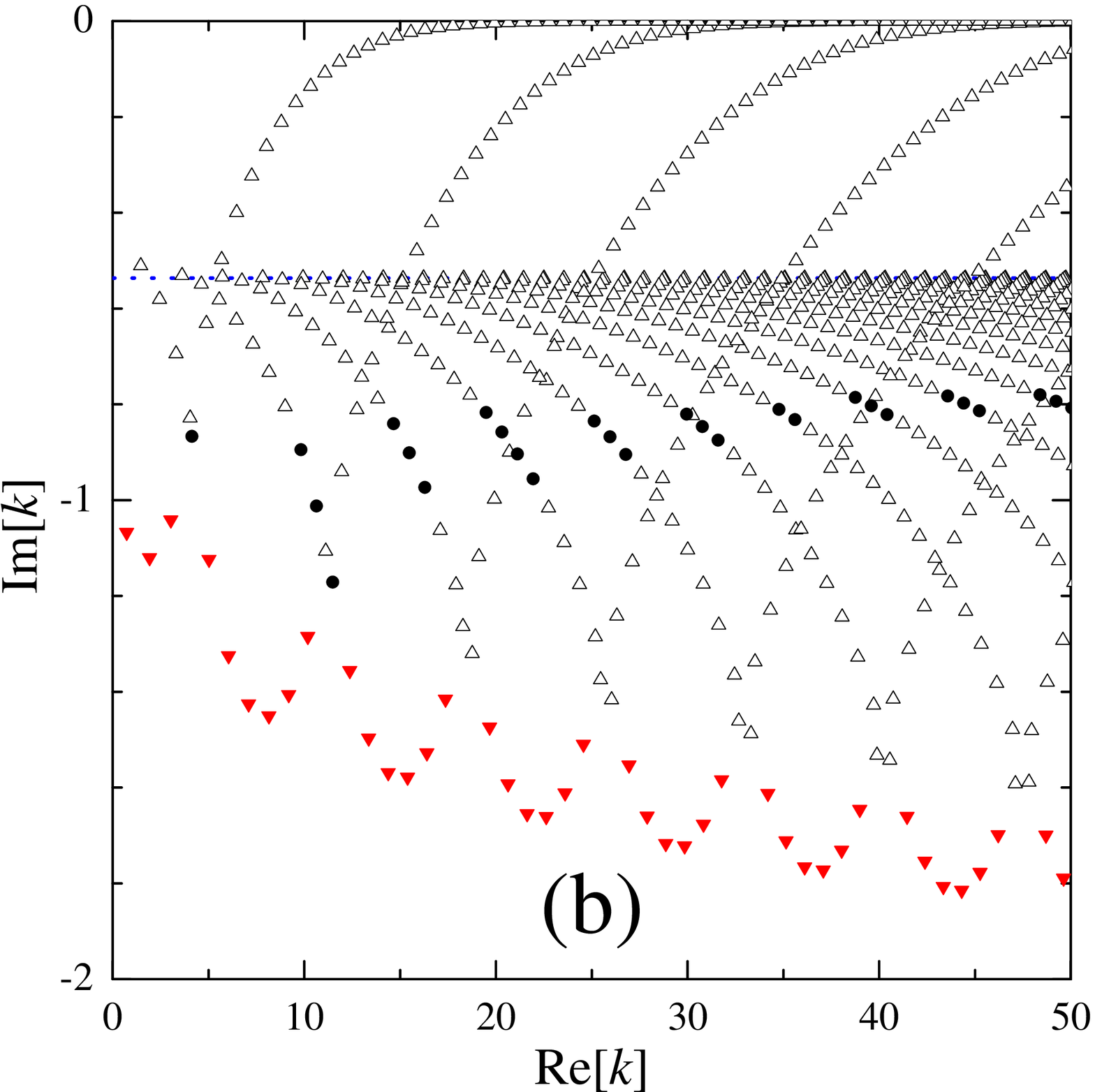}
\par\end{centering}

\caption{(Color online) Resonance positions at $\Omega=0.667$ for TE polarization.
Inverted triangles are $\text{o}(m,\left[\frac{m+1}{2}\right])$
and the others are $\text{i}(m,l)$ decided by considering ARC. Circles
are $\text{i}(m,l)$ changed from $\text{o}_{0}(m,\left[\frac{m+1}{2}\right])$.
 Dashed, dotted, and dot-dashed lines in (a) are $\text{Re}[k]=m\Omega$,
$\text{Re}[k]=m$, and $k_{B}$, respectively. The inset in (a)
is for $\Omega=0.5$. Blue dotted line in (b) is $\gamma$. \label{fig:ResPos_n1.5_ARC}}

\end{figure}

The dynamics from the numerical result and the toy model is explained
in terms that the coupling between cavity itself and environment
becomes strong as openness increases, and then the IRs are exquisitely
affected by low-leaky ORs in large opening regime.   Hence, a
care should be taken to the identification of MDRs according to the
history of ARCs.  On a resonance position plot in the large opening
regime, various IRs can coexist such as once and twice swapping resonances
 after strong ARCs, those superposed near weak ARCs and strong
ARCs, and the uncoupled ones. Figure \ref{fig:ResPos_n1.5_ARC} shows
the resonance positions in consideration of the ARCs. We included
only the IRs and the low-leaky ORs for the sake of convenience.
The dotted and the dashed lines in Fig. \ref{fig:ResPos_n1.5_ARC}(a)
are $\text{Re}[k]=m$ and $\text{Re}[k]=m\Omega$, respectively. In
the effective potential analogy, $\text{Re}[k]=m$ corresponds to
the top of the effective potential barrier, that is, the critical
incident angle for IRs and $\text{Re}[k]=m\Omega$ corresponds to
the bottom of the barrier, that is, the incident angle $\theta=\pi/2$.
Effective potential analogy is successfully applied to schematic description
of the properties for IRs. For this reason, it is often used on the
research about the resonances in dielectric disk. We, however, note
that the effective potential analogy has some insufficient parts yet
on applying to ORs\cite{Cho}. For a given $m$ in Fig. \ref{fig:ResPos_n1.5_ARC}(a),
a circle is $\text{i}(m,l)$ changed from $\text{o}_{0}(m,\left[\frac{m+1}{2}\right])$
by the first strong ARC  and an inverted triangle is $\text{o}(m,\left[\frac{m+1}{2}\right])$
changed from $\text{i}_{0}(m,l)$ by the final strong ARC. The triangles
between the circle and the inverted triangle are $\text{i}(m,l-1)$
changed from $\text{i}_{0}(m,l)$ by two consecutive strong ARCs.
 If we are able to get anyhow a correct position of $\text{o}(m,\left[\frac{m+1}{2}\right])$,
it is natural that $\text{i}_{0}(m,l)$, whose $\text{Re}[k_{r}]$s
are positioned between $\text{o}_{0}(m,\left[\frac{m+1}{2}\right])$
and $\text{o}(m,\left[\frac{m+1}{2}\right])$, are $\text{i}(m,l-1)$
\cite{Dettmann}, because  $\text{o}(m,\left[\frac{m+1}{2}\right])$
must be excluded from the ordering of $l$ for $\text{i}(m,l)$.

Here, we newly use a line number for ORs defined as $q=\left[\frac{m+1}{2}\right]-l$,
$(q=0,1,2,\cdots)$ and, then, the low-leaky ORs are on the line
of $q=0$. The individual line of $q$ except zero is generally parallel
to the line of $\text{Re}[k]=m$. However, the line of $q=0$ has
 a different slope. It surprisingly corresponds to the Brewster
angle in large opening regime. In the viewpoint of semiclassics,
the incident angle for an IR can be estimated by  $\sin\theta=m\Omega/\text{Re}[k]$
and the Brewster angle is represented as $\theta_{B}=\arctan\left(\Omega\right)$.
Consequently, one can obtain $\text{Re}[k]$ relevant to the Brewster
angle, that is, $k_{B}=m\sqrt{1+\Omega^{2}}.$ In Fig. \ref{fig:ResPos_n1.5_ARC}(a)
and inset,  $\text{Re}[k_{r}]$s of low-leaky ORs are well-aligned
on the $k_{B}$ line, whether they have experienced strong ARC or
not. Also, by strong ARC in the opening regime where IRs and low-leaky
ORs are coexistent in the same region of $k$ plane, $\text{Im}[k_{r}]$s
of low-leaky ORs settle down in the bottom region of the IR group
as shown in Fig \ref{fig:ResPos_n1.5_ARC}(b).

\begin{figure}
\begin{centering}
\includegraphics[width=8.5cm]{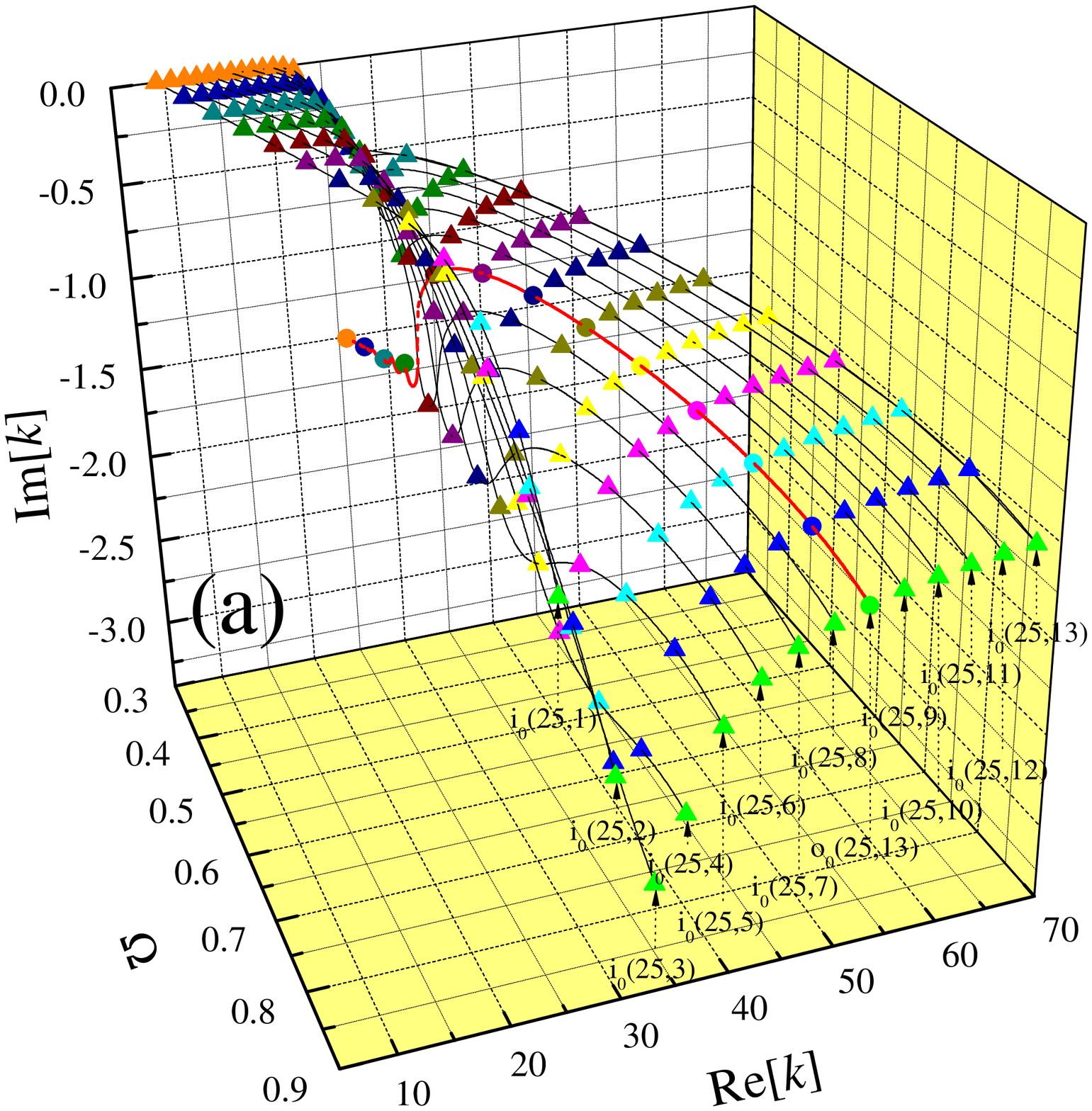}
\par\end{centering}

\begin{centering}
\includegraphics[width=8.5cm]{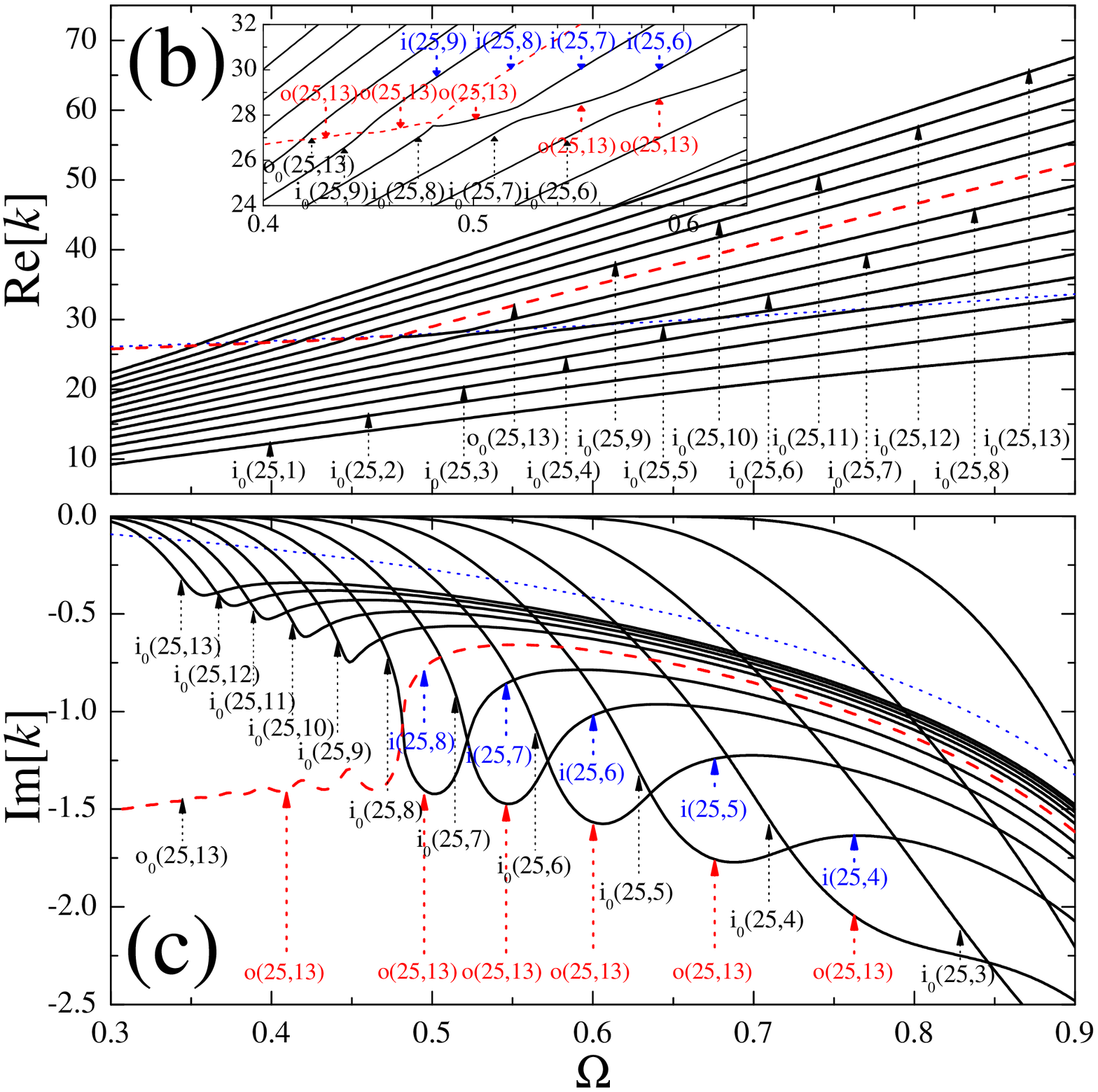}
\par\end{centering}

\caption{(Color online) Dynamics of MDR with $m=25$ by openness. Solid lines
and a dashed line are $\text{i}_{0}(25,l\leq13)$ and $\text{o}_{0}(25,13)$,
respectively. Dotted line in (b) and (c) is $k_{B}$ and $\gamma$,
respectively. \label{fig:res_dynamics}}

\end{figure}

Figure \ref{fig:res_dynamics}(a) is  the three-dimensional representation
of the dynamics of $\text{i}_{0}(25,l\leq13)$ and $\text{o}_{0}(25,13)$
in $\text{Re}[k_{r}]$, $\text{Im}[k_{r}]$ and $\Omega$ space. For
more detailed depiction and notation of reference marks, we projectively
plot the dynamics of $\text{Re}[k_{r}]$ and $\text{Im}[k_{r}]$ in
Fig. \ref{fig:res_dynamics}(b) and Fig. \ref{fig:res_dynamics}(c),
respectively. The dynamics of $\text{Re}[k_{r}]$ for IRs and ORs
in TM polarization exhibits  almost linear and constant behaviors,
respectively \cite{Dubertrand,Bogomolny,Noeckel}. On the other hand,
in TE case, the low-leaky ORs breaks the monotonic behavior of
IRs as shown in Fig. \ref{fig:res_dynamics}(b) through the ARCs.
 As $\Omega$ increases, $\text{o}_{0}(25,13)$ interacts with $\text{i}_{0}(25,l)$,
where $13\geq l\geq9$, sequentially through a series of weak ARCs
before it interacts with $\text{i}_{0}(25,8)$ through the first strong
ARC at $\Omega\simeq0.481$, i.e., there is a transition from weak
ARC to strong ARC. In the large opening regime, the point of strong
ARC satisfies $k_{B}$ for given $m$ and $\Omega$, since $\text{Re}[k_{r}]$s
of IRs increase while $\text{Re}[k_{r}]$s of low-leaky ORs tend
to maintain near $k_{B}$. After the first strong ARC, $\text{Re}[k_{r}]$
of $\text{i}_{0}(25,8)$ stays near $k_{B}$ because the resonance
turns to $\text{o}(25,13)$ by the exchange of their identities. 
The strong ARCs between $\text{o}(25,13)$ and $\text{i}_{0}(25,l)$,
where $l\leqq8$, sequentially occur at $\text{Re}[k_{r}]\simeq k_{B}$
in the region of $\Omega\geqq0.481$ until $\Omega$ becomes unity,
where all resonances merge into the continuum.

The dynamics of $\text{Im}[k_{r}]$ shows more interesting behavior.
$\text{Im}[k_{r}]$ for a resonance is directly connected to the lifetime
($\tau=-1/2c\text{Im}[k]$) and $Q$-factor ($Q=-\text{Re}[k]/2\text{Im}[k]$).
 On the space of $\text{Im}[k_{r}]$ vs $\Omega$ as shown in Fig.
\ref{fig:res_dynamics}(c), IR and low-leaky OR approach each other
near the point of ARC. Since the low-leaky OR has lower $Q$  than
the IRs for the same $m$,  $\left|\text{Im}[k_{r}]\right|$ of
$\text{i}(25,l)$ increases  by the interaction with $\text{o}(25,13)$.
 Moreover, the increased $\left|\text{Im}[k_{r}]\right|$ after ARC
is always below the bound line $\gamma$ due to the aftermath of the
ARC. Hence, at a fixed value of $\Omega$, the IRs positioned below
the $\gamma$ line are the resonances influenced by the ARCs. This
is the reason why the IRs for TE polarization have generally lower
$Q$ than the corresponding modes for TM case \cite{Dubertrand,Bogomolny,Cho,Dettmann,Ryu}
and their imaginary part falls below the $\gamma$ line in the ordinary
experimental regime of $\Omega$ as shown in Fig \ref{fig:ResPos_n1.5_ARC}(b). 

Also, the dynamics gives  a clear explanation for the breaking of
the decreasing order of $Q$-factor in accordance with the increasing
order of $l$ near the Brewster angle. In a local region of complex
$k$ space under experimentally feasible circumstances, the grouping
of resonances by the order of $Q$  makes it possible for us  to
predict the characteristics of  the resonances. If the effects of
ARC are practically zero, the IR with higher $l$ has generally
low $Q$  according to the effective potential analogy since a resonance
with higher energy is less trapped in the effective potential \cite{Johnson,Cho,thesis_Noeckel,thesis_Hentschel}.
In TE case, however, the order of $Q$ of IRs is reversed from a
certain value of $l$ due to ARCs with ORs as shown in Fig. \ref{fig:res_dynamics}(c).
For instance, at $\Omega=0.5$, the order of $Q$ corresponds to the
order of $l$ in TM case as $Q_{\text{i}(25,9)}=85.876>Q_{\text{i}(25,10)}=83.253>Q_{\text{i}(25,11)}=82.528$.
On the other hand, in TE case, the order is reversed as $Q_{\text{i}(25,9)}=27.116<Q_{\text{i}(25,10)}=33.120<Q_{\text{i}(25,11)}=38.377$.
As $\Omega$ increases, this phenomenon is extended to the resonances
with lower $l$ due to the increase of the coupling strength.

Such phenomena, only occur for TE polarization case, indicate that
the existence of Brewster angle strongly activate the interactions
between IRs and ORs as $\Omega$ increases. In the dielectric disk
for TE polarization, the IRs are sensitively affected by Brewster
angle for the internal cavity boundary.  So, the IRs whose $\text{Re}[k_{r}]$
is near $k_{B}$ have larger leakage ($\left|\text{Im}[k]\right|$)
than the others \cite{Ryu}. i.e., the more $\text{Re}[k_{r}]$ of
IRs approaches to $k_{B}$, the more the leakage of the IR increases.
An IR whose leakage is increased near the Brewster angle by the change
of openness, as $\Omega$ increases, approaches to a low-leaky OR
in complex $k$ space. Then, the coupling with the low-leaky OR becomes
relatively strong and the interaction between them is more activated.
Such ARC for each IR occurs sequentially at other points of $\Omega$.
Moreover, since the coupling strength becomes strong as $\Omega$
increases, the transition to strong ARC from weak ARC occurs based
on an IR with specific value of $l$ as shown in Fig. \ref{fig:res_dynamics}.

\section{Summary}

Summarizing, we have demonstrated that the MDR dynamics by openness
in a dielectric circular microdisk for TE polarization shows ARCs
over two types of resonances due to the coupling with environment.
The low-leaky outer resonances interact  pronouncedly with the inner
resonances through the opening channel near the Brewster angle. 
As a result, in the opening regime relevant to experiments, the resonance
classification between inner and outer resonances is more complicated
and the change of leakage for inner resonances occurs. The reverse
order of $Q$-factors of inner resonances which is discrepant from
the effective potential analogy can be understood in terms of the
ARCs between inner and outer resonances.   This study will be useful
as a basis to analyze intricate resonance dynamics in deformed dielectric
microcavities.

\section*{ACKNOWLEDGMENTS}

We would like to thank I. Kim, J.-W. Ryu, S.-Y. Lee, and T. Harayama
for helpful discussions. This work was supported by Acceleration Research
(Center for Quantum Chaos Applications) of MEST/KOSEF.


\begin{thebibliography}{26}
\bibitem{Chang} R.K. Chang and A.J. Campillo (Eds.), Optical Processes
in Microcavities (World Scientific, Singapore, 1996).

\bibitem{Johnson} B.R. Johnson, J. Opt. Soc. Am. A \textbf{10}, 343
(1993).

\bibitem{Noeckel} J.U. N{\"o}ckel and A.D. Stone, Nature (London)
\textbf{385}, 45 (1997).

\bibitem{Dubertrand} R. Dubertrand, E. Bogomolny, N. Djellali, M.
Lebental, and C. Schmit, Phys. Rev. A \textbf{77}, 013804 (2008),

\bibitem{Bogomolny} E. Bogomolny, R. Dubertrand, and C. Schmit, Phys.
Rev. E \textbf{78}, 056202 (2008).

\bibitem{Harayama} S. Tasaki, T. Harayama, and A. Shudo, Phys. Rev.
E \textbf{56}, R13 (1997).

\bibitem{Dettmann} C.P. Dettmann, G.V. Morozov, M. Sieber, and H.
Waalkens, Europhys. Lett. \textbf{87}, 34003 (2009).

\bibitem{Cho} J. Cho, I. Kim, S. Rim, G.-S. Yim, and C.-M. Kim, Phys.
Lett. A \textbf{374}, 1893 (2010).

\bibitem{Stoeckmann} H.-J. St{\"o}ckmann, Quantum Chaos (Cambridge
University Press, Cambridge, England, 2000).

\bibitem{ARC_open_Hamiltonian} W.D. Heiss, Phys. Rev. E \textbf{61},
929 (2000), A.I. Magunov et al., Physica E \textbf{9}, 474 (2001).

\bibitem{ARC_chaotic_theory_Wiersig_Hentschel} J. Wiersig and M.
Hentschel, Phys. Rev. A \textbf{73}, 031802(R) (2006).

\bibitem{ARC_chaotic_theory_SYLee} S.-Y. Lee, J.-W. Ryu, J.-B. Shim,
S.-B. Lee, S.W. Kim, and K. An, Phys. Rev. A \textbf{78}, 015805 (2008).

\bibitem{ARC_chaotic_theory_Wiersig_SWKim_Hentschel} J. Wiersig,
S.W. Kim, and M. Hentschel, Phys. Rev. A \textbf{78}, 053809 (2008).

\bibitem{ARC_chaotic_theory_Ryu} J.-W. Ryu, S.-Y. Lee, and S.W. Kim,
Phys. Rev. A \textbf{79}, 053858 (2009).

\bibitem{ARC_chaotic_theory_Lizuain} I. Lizuain, E. Hern{\'a}ndez-Concepci{\'o}n,
and J.G. Muga, Phys. Rev. A \textbf{79}, 065602 (2009).

\bibitem{ARC_chaotic_theory_Poli} C. Poli, B. Dietz, O. Legrand,
F. Mortessagne, and A. Richter, Phys. Rev. E \textbf{80}, 035204(R)
(2009).

\bibitem{ARC_chaotic_exp_Dietz} B. Dietz, A. Heine, A. Richter, O.
Bohigas, and P. Leboeuf, Phys. Rev. E \textbf{73}, 035201(R) (2006)

\bibitem{ARC_chaotic_exp_SBLee} S.-B. Lee, J. Yang, S. Moon, S.-Y.
Lee, J.-B. Shim, S.W. Kim, J.-H. Lee, and K. An, Phys. Rev. A \textbf{80},
011802(R) (2009).

\bibitem{ARC_chaotic_exp_SBLee2} S.-B. Lee, J. Yang, S. Moon, S.-Y.
Lee, J.-B. Shim, S.W. Kim, J.-H. Lee, and K. An, Phys. Rev. Lett.
\textbf{103}, 134101 (2009).

\bibitem{ARC_integrable_Wiersig} J. Wiersig, Phys. Rev. Lett. \textbf{97},
253901 (2006).

\bibitem{ARC_integrable_Unterhinninghofen} J. Unterhinninghofen,
J. Wiersig, and M. Hentschel, Phys. Rev. E \textbf{78}, 016201 (2008).

\bibitem{Creagh} S.C. Creagh, Phys. Rev. Lett. \textbf{98}, 153901
(2007).

\bibitem{Lee} J. Lee, S. Rim, J. Cho, and C.-M. Kim, Phys. Rev. Lett.
\textbf{101}, 064101 (2008).

\bibitem{thesis_Noeckel} J.U. N{\"o}ckel, Ph.D. Thesis, Yale University,
1997.

\bibitem{thesis_Hentschel} M. Hentschel, Ph.D. Thesis, Max Planck
Institute for the Physics of Complex Systems, 2002.

\bibitem{Ryu} J.-W. Ryu, S. Rim, Y.-J. Park, C.-M. Kim, and S.-Y.
Lee, Phys. Lett. A \textbf{372}, 3531 (2008).
\end{thebibliography}
\end{document}